# *In situ* and *operando* characterisation techniques for solid oxide electrochemical cells: Recent advances

Alexander Stangl, David Muñoz-Rojas and Mónica Burriel*

Univ. Grenoble Alpes, CNRS, Grenoble INP*, LMGP, 38000 Grenoble, France

* Institute of Engineering Univ. Grenoble Alpes

## Abstract:

Oxygen activity and surface stability are two key parameters in the search for advanced materials for intermediate temperature solid oxide electrochemical cells, as overall device performance depends critically on them. In particular *in situ* and *operando* characterisation techniques have accelerated the understanding of degradation processes and the identification of active sites, motivating the design and synthesis of improved, nanoengineered materials. In this short topical review we report on the latest developments of various sophisticated *in situ* and *operando* characterization techniques, including Transmission and Scanning Electron Microscopy (TEM and SEM), surface-enhanced Raman spectroscopy (SERS), Electrochemical Impedance Spectroscopy (EIS), X-ray Diffraction (XRD) and synchrotron based X-ray photoelectron spectroscopy (XPS) and X-ray absorption spectroscopy (XAS), among others. We focus on their use in three emerging topics, namely: (i) the analysis of general electrochemical reactions and the surface defect chemistry of electrode materials; (ii) the evolution of electrode surfaces achieved by nanoparticle exsolution for enhanced oxygen activity and (iii) the study of surface degradation caused by Sr segregation, leading to reduced durability. For each of these topics we highlight the most remarkable examples recently published. We anticipate that ongoing improvements in the characterisation techniques and especially a complementary use of them by multimodal approaches will lead to improved knowledge of *operando* processes, hence allowing a significant advancement in cell performance in the near future.

## Introduction:

Anthropogenic global warming is one of the biggest threats to modern human society [1] and requires a transition from fossil fuel based to sustainable energy systems, including renewable energy harvesting and high efficiency energy conversion for storage. Energy conversion based on solid oxide cells (SOCs) provides key features as high efficiency, low production and operating cost, durability and potential zero carbon emission [2]. However, the optimization of the different parts of SOCs in realistic operation conditions is yet to be achieved. This would involve analysis and long-term testing in ambient air, containing pollutants such as nitrogen, carbon (such as $CO_2$ and hydrocarbons) and sulfur oxides and water vapor. These pollutants are known to impact the chemical composition of the surface and hence may strongly modify exchange kinetics.

*In situ* and *operando* characterisation techniques are becoming key for the advancement in the knowledge and understanding of the complicated electrochemical processes which take place in SOC cells [3]–[5]. These techniques are now able to answer questions, which previously could only be addressed by indirect electrochemical measurements (e.g. impedance spectroscopy analysis) or by *post mortem ex situ* analysis using complementary structural, morphological and physio-chemical characterisation techniques [6]. Unfortunately *post mortem ex situ* analysis is based on one or several "camera shots" taken to a sample at room temperature, after having been exposed to a specific environment (temperature, oxygen partial pressure, anodic or cathodic polarization, etc.) for a particular amount of time. It is clear that obtaining *in situ*, "live" measurements, i.e. the "whole movie" of the electrochemical processes with time-resolved resolution, provides much richer information and can help directly link structural and chemical changes at conditions similar to operation (e.g. high temperature, atmospheric pressure) with the electrochemical response of the system. While *in situ* measurements are often limited to an individual cell component, i.e. the cathode, in specific environments, *operando* (Latin for "working") techniques monitor a device during operation. This is a highly sophisticated challenge, especially for SOCs, as it requires an elaborate cell design including heating systems, gas connections and fuel supply and firm isolation of the electrode atmospheres, in addition to experimental probes and instrumentation.

This topical review aims to provide a snapshot of the most recent developments in this field, now that improvements in *in situ* and *operando* measurement environments are step by step approaching realistic operation conditions. Although it focuses on SOCs applications, it is noteworthy mentioning that similar advances are being undertaken in other types of electrochemical and optoelectronic devices such as batteries and solar cells [7]–[11].

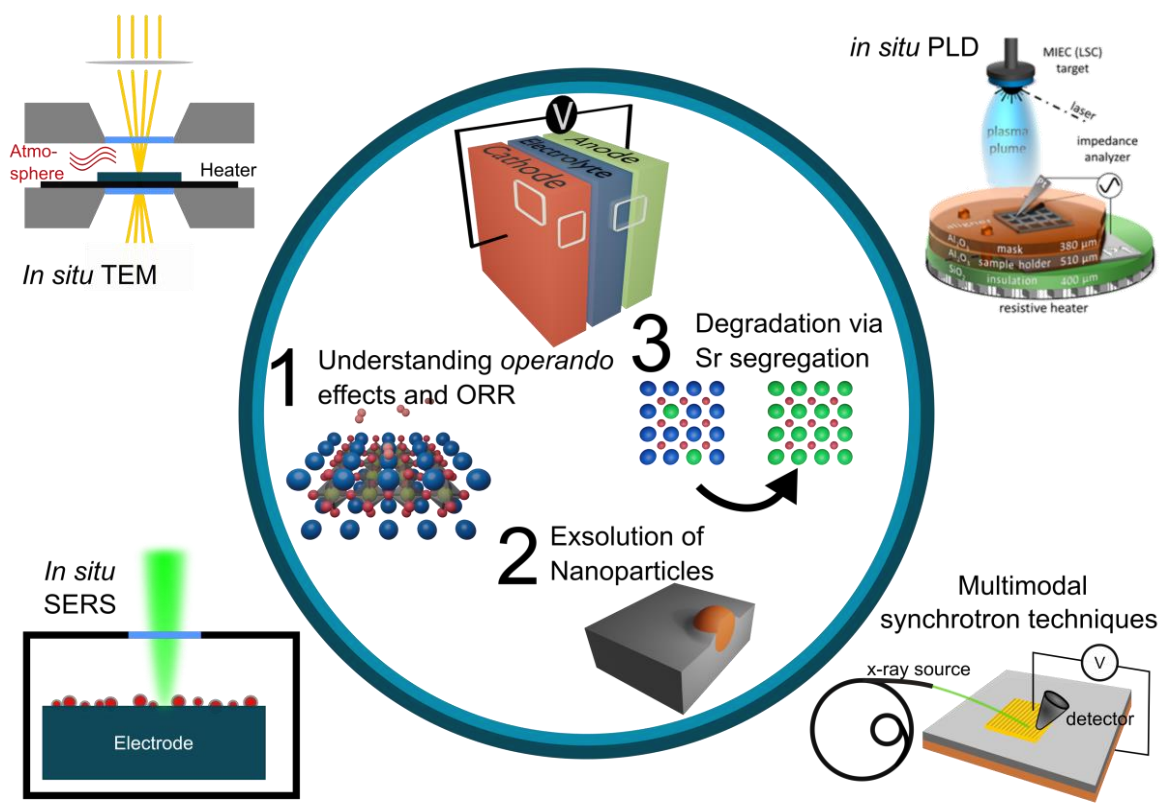

*Figure 1:**In situ and operando characterisation for improved cell performance.** The three main topics of interest as well as some of the covered in situ and operando techniques are schematically represented. IPLD schematic reproduced under Creative Commons terms (CC-BY) from* [12]*.*

Targeting the main source of polarisation losses is critical for achieving real breakthroughs of the SOC technologies. Here we focus on *in situ* and *operando* characterisation of electrode materials, as their contribution typically governs the overall cell resistance, effectively reducing the possible power output of a fuel cell. We centre our attention on some of the most actively studied materials for intermediate temperature Solid Oxide Fuel Cells (SOFCs)[13], [14] and solid oxide electrolysis cells (SOECs)[15], and mainly on the oxides used as oxygen electrode. These are generally simple perovskites [16], including model systems such as A-site Sr doped $LaCoO_{3-\delta}$ (LSC), as well as other layered oxides, such as double perovskites and the Ruddlesden-Popper phases [17]. The techniques discussed, comprising advanced laboratory scale techniques such as Transmission Electron Microscopy (TEM) and surface enhanced Raman spectroscopy, as well as many different synchrotron based X-ray techniques, are summarised in Figure 1. We first focus on the analysis of elementary processes, e.g. on the physio- and electrochemical reactions, and on the essential understanding of local defect chemistry. At the same time we briefly discuss the specific features of each technique presented together with the main results and information obtained in the works reviewed. Then we report on the most recent approaches to study the mechanisms of exsolution of nanoparticles for enhanced cathode performance. Finally we discuss strontium segregation, as a key degradation issue in perovskite-based cathodes, focussing on its origin and impact on electrode activity. We wrap the review with some prospects on the future use of these techniques, giving some hints on how they can contribute to the understanding of the complex temperature-, field-, chemistry- and time-dependant electrochemical processes taking place. We think cross fertilization between the techniques applied to the different types of electrochemical devices should pave the way for further advancement.

# Electrochemical reactions and defect chemistry analysis of electrode materials and interfaces

Most of the work carried out nowadays to improve the performance of solid oxide electrochemical cells and systems is focused on lowering the polarization resistance of the electrodes. To do so it is first important to understand the defect chemistry of the materials, the electrochemical processes taking place in the cell (e.g. oxygen reduction, fuel oxidation, ion transport) and the rate determining steps (RDS) governing under operating conditions. This section highlights some of the most relevant literature examples published in the last years based on the use of *in situ* and *operando* characterization techniques for the general analysis of electrochemical reactions and for the study of the surface defect chemistry of electrode materials. The articles presented have been selected as illustrating examples of each technique and their potential use and focus on state-of-the art relevant electrode materials such as LSC, $La_2NiO_{4+\delta}$ and $Ce_{1-x}Gd_xO_{2-x/2}$ (CGO) using standard electrolytes such as yttria-stabilized zirconia (YSZ) and CGO. The examples presented move from more accessible laboratory based advanced *in situ* characterization techniques to Synchrotron based X-ray analysis including both *in situ* and *operando* measurements.

## Laboratory based techniques
Raman spectroscopy constitutes a very simple and effective tool which has been extensively used *in situ* and *ex situ* to probe surface species such as oxygen, sulfur, hydrocarbons and water [6] and to probe structural changes upon oxidation or reduction, resulting in changes in the vibration modes of the electrode materials' structure. For further details on this technique the reader is addressed to the following comprehensive reviews [3], [6], [18]. Being one of the limitations of Raman spectroscopy its lack of sensitivity to many surface species, enhanced Raman spectroscopy (SERS) has aroused as a promising solution. SERS uses nanoparticles of metals such as Au or Ag to amplify the signal of the molecules in their vicinity. The use of $SiO_2$ shell isolated Ag nanoparticles enabled the applicability of SERS at high temperatures (~400 ºC) boosting its sensitivity when compared to conventional Raman spectroscopy. This new methodology was used for the first time for SOFC materials by Liu's group [19], showing its applicability both for studying the coking (shown in Figure 2(a-c)) and carbon removal on Ni (SOFC anode) and the accumulation and depletion of adsorbed oxygen in ceria under oxidizing and reducing conditions, respectively.

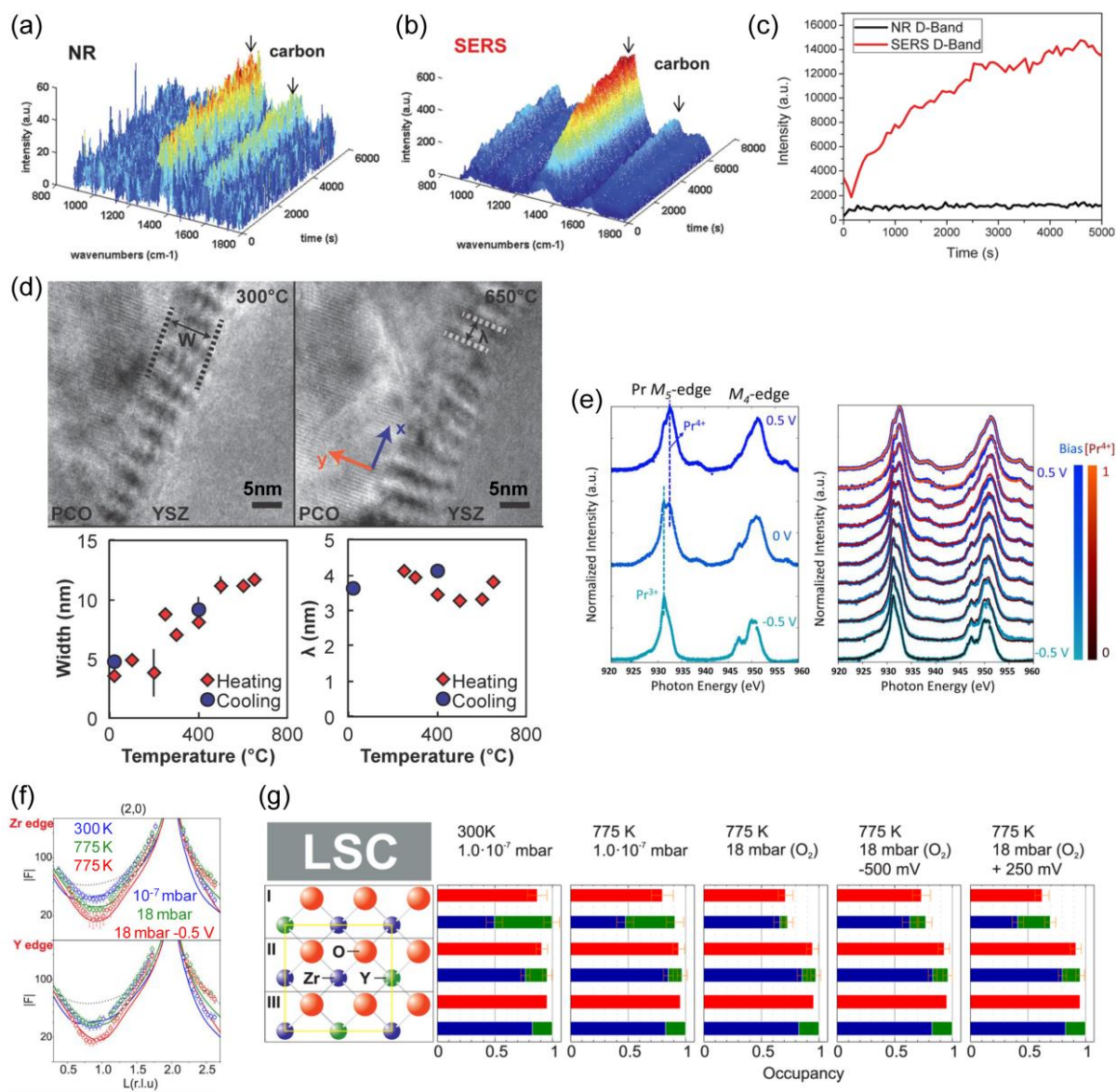

*Figure 2: **Study of electrochemical reactions and defect chemistry using advanced characterisation techniques.** In situ analysis of carbon coking on Ni via **(a)** normal Raman spectroscopy and **(b)** SERS with Ag/SiO₂ core shell nanoparticles. The integrated intensities of carbon D-band mode at 1350 cm$^{-1}$ from **(a)** and **(b)** are shown in **(c)** [19]. © 2014, The Royal Society of Chemistry. **(d)** In situ phase contrast TEM images of $Pr_xCe_{1-x}O_2$-YSZ interface at 300°C and 650°C (upper panels) and temperature evolution of interfacial fringe width, W and periodicity, λ (lower panel) [20]. © 2018, American Chemical Society. **(e)** Operando X-ray adsorption spectra of PCO at 450°C and pO₂= 200 mTorr at different polarisation conditions [21]. © 2018, American Chemical Society. (f) Operando anomalous surface X-ray diffraction study of the YSZ/LSC interface X-ray crystal truncation rod scattering and (g) deduced model of occupancy of the first three atomic YSZ layers below the LSC electrode. Reproduced under Creative Commons (CC-BY) terms from [22]. © 2016, American Chemical Society.*

*In situ* transmission electron microscopy (TEM) is another very powerful, yet more complex and expensive characterization technique, widely used to visualize electrochemical processes at the atomic scale in real time. It has been used for a broad range of energy materials and devices including batteries, catalysts, fuel cells and solar cells. Examples of recent progress and specific setups developed for the *in situ* measurements can be found in the following excellent recent review articles [23], [24]. While *in situ* TEM had opened a window into the analysis under different atmospheres, currently accessible pressure ranges are still far from real operation conditions and advances are needed to extend the applicability of this technique.

A beautiful example of the use of *in situ* TEM combined with scanning transmission electron microscopy (STEM) and electron energy loss spectroscopy (EELS) was reported for the study $Pr_xCe_{1-x}O_{2-\delta}$ (PCO) deposited as a thin film on YSZ [20]. By performing a series of consecutive experiments at three different temperatures (room temperature, 300 °C and 650°C) in oxidizing and reducing conditions, the authors were able to characterize chemical strains and changes in oxidation state in PCO lamella cross sections. Interestingly, changes in structural defects were also observed; while periodic arrays of strain fields, consistent with misfit dislocation features, changed their size during thermochemical expansion (Figure 2(d)), threading dislocations, able to trap reduced cations, appeared upon reoxidation of the sample. In addition, the sample was observed to "breathe" oxygen out of the lattice with an anisotropic chemical expansion which varied with the distance from the interface. A very interesting method developed by Fleig's group consists in performing *in situ* impedance spectroscopy during pulsed laser deposition (IPLD), as illustrated in the sketch shown in Figure 1 [12]. LSC was used as model material in this first study, and was deposited on top of an electrolyte material with a porous LSC counter electrode at the bottom and a Pt grid on top. With this smart configuration the LSC film can be electrochemically characterized by electrochemical impedance spectroscopy (EIS) as it is grown, effectively obtaining the oxygen exchange kinetics and defect chemistry of the as-deposited samples as a function of film thickness. The authors reported surprisingly repeatable results for the LSC films with a much higher electrochemical activity and higher reproducibility than films measured in standard equipment, which typically show a large scatter in polarization resistance.

Surfaces are utterly delicate and surface activity can be drastically varied upon changes in atmosphere and temperature. Thus, *in situ* studies during or immediately after the deposition are highly attractive, as they provide information about "fresh", as-deposited surfaces, which in the case of LSC were shown to be electrochemically extremely active. Identifying and preserving these beneficial pristine properties of a material will be the next step on the route to materials by design.

## Synchrotron based X-ray techniques

A large number of synchrotron based materials characterization techniques exist, being X-ray absorption spectroscopy (XAS), X-ray diffraction (XRD), ambient pressure XPS (AP-XPS) and X-ray Raman spectroscopy (XRS) the most relevant for the characterization of electrochemical oxide materials. Remarkably, when moving from a laboratory X-ray source to a synchrotron source the brilliance can be increased by up to a billion times, explaining the attractiveness of these techniques which allow fast data acquisition, highly improved quality, and measurements that otherwise would not have been attainable (at ambient pressure, reaching deep interfaces, etc). Furthermore, in some beamlines several complementary measurement techniques can be simultaneously combined in a single experiment which, when used together with dedicated electrochemical cells, can provide time-resolved structural, chemical and electrochemical information as the materials evolve or as the device operates.

In other cases measurements on the same samples are carried out on different beamlines following similar *in situ* treatments, as in the case of the study of Ni doped $La_{0.6}Sr_{0.4}Co_{0.2}Fe_{0.8}O_{3-\delta}$ (LSCF) cathodes as a function of temperature (up to 950 °C) [25]. The authors combined XAS (X-ray absorption near edge spectroscopy (XANES) and extended X-ray absorption fine structure (EXAFS)) measurements at the Fe, Co, and Ni K-edges with X-ray Raman scattering (XRS) measurements at the O K-edge at two different beamlines. The information obtained, together with impedance measurements of symmetrical cells with electrodes of the same composition, revealed the effect of Ni doping on the electrochemical performance and electronic structure of LSCF cathodes. The improved performance obtained by Ni doping is explained by the role of the $Ni^{3+}$ cation hindering the formation of oxygen vacancies near Co, thus stabilizing the Co higher oxidation state. Furthermore, Ni helps retain the hybridization between the O 2p and the transition metal 3d orbitals, and thus hinders the modification of the perovskite octahedra maintaining the redox properties over a wide temperature range in oxidizing conditions.

XAS measurements at the Pr $M_{4,5}$-edge, here combined with AP-XPS in the same beamline, were also used to study the surface defect chemistry and electronic structure of $Pr_{0.1}Ce_{0.9}O_{2-\delta}$ (PCO) by Yildiz's group who carried out *operando* measurements on a full electrochemical cell [21]. It should be mentioned that *operando* AP-XPS at elevated temperatures had previously been reported by Chueh et al. [26] and Feng et al. [27] for Sm-doped ceria in a reducing $H_2/H_2O$ atmosphere. By studying Pr-doped ceria, with a relatively high ionic and electronic conductivity, as model material Lu *et al*. [21] were able to extend the atmosphere to a high-$pO_2$ oxidizing regime and simulate the operating conditions of SOFC cathodes (or SOEC anodes). Both cathodic and anodic electrochemical potential were applied in voltage steps to the PCO electrode (at 450 °C and 200 mTorr), at each of which the concentration of reduced $Pr^{3+}$ could be determined from the XAS spectra, reported in Figure 2(e). The surface defect concentration could then be plotted as a function of a wide range of effective $pO_2$ (using the Nernst equation), e.g. $\log[Pr'_{Ce}]$ vs $\log pO_2$, indicating a much shallower slope for the PCO surface compared to the values predicted from bulk defect chemistry. In addition, by performing resonant XPS, the energy position of the Pr defect states in the electronic structure was also resolved.

Chueh's group has also just reported another study on PCO using *operando* AP-XPS and ambient pressure XAS under polarization in full electrochemical cells (illustration of the setup used shown in Figure 1) [28], which aim was to determine the RDS for the oxygen incorporation in PCO. They successfully determined the most probable RDS as the dissociation of neutral molecular oxygen adsorbate from over one hundred possibilities. This remarkable result was achieved by measuring the current density-overpotential curves at high temperatures while controlling the oxygen gas partial pressures and the oxygen activity in the solid. The authors quantified the chemical and electrostatic driving forces by measuring the O 1s binding energy by *operando* AP-XPS as a function of applied voltage, while the Pr M-edge measurement gave information on the concentrations of $Pr^{+3}$ and $Pr^{4+}$ cation concentrations as a function of overpotential. The importance of this elegant study comes, not only from the determination of the RDS for this electrode, but more widely from the development of an experimental and analysis framework that can be applied in the future to determine the oxygen incorporation RDS in many other electrochemical oxides, as for identifying the RDS of other reactions in systems such as polymer fuel cells or lithium batteries.

Another exciting study focused on the RDS determination by *operando* XAS was carried out by Amezawa's group on fuel cells using $La_2NiO_{4+\delta}$ dense-film electrodes [29]. In this work soft X-rays were used, allowing the measurement of both the O K-edge and the Ni L-edge X-ray absorption spectra at 500 °C, at several oxygen partial pressures and under polarization. While only small changes were observed for the Ni L-edge, by changing these 2 parameters considerable spectral changes were observed in the O K-edge spectra. The pre-edge, which reflects the unoccupied partial density of states of Ni 3d - O 2p hybridization, increased and decreased upon cathodic and anodic polarization, respectively, while an additional feature appeared at high oxygen partial pressure due to the adsorption of oxygen molecules in the gas phase. A surface process was found to be the RDS, with a $pO_2$ dependence that suggests either a dissociative adsorption of oxygen molecules or the charge transfer as RDS. However, in this case further characterization would be required to univocally elucidate the reaction mechanisms.

In the following example near ambient pressure X-ray photoelectron spectroscopy (NAP-XPS) and impedance spectroscopy were used simultaneously to investigate acceptor-doped perovskite-type $La_{0.6}Sr_{0.4}CoO_{3-\delta}$ (LSC), $La_{0.6}Sr_{0.4}FeO_{3-\delta}$ (LSF), and $SrTi_{0.7}Fe_{0.3}O_{3-\delta}$ (STF) thin film model electrodes [30] revealing numerous details of the oxides defect chemistry. For all three compositions additional species of strontium and oxygen were measured in oxidizing conditions, while switching between oxidizing and reducing conditions resulted in a shift of the apparent binding energy. In the same way, the binding energy strongly depended on the applied electrochemical polarization. Both shifts were explained by the correlation of Fermi level and the chemical potential of oxygen at given conditions. Finally, differences in the valence band structure could be related to different electronic conduction mechanisms: polaron-type electronic conduction for LSF and STF and the metal-like conduction for LSC.

Electrostatic effects, due to the formation of interfacial dipoles, impact the rate of the ion insertion reaction at the gas/MIEC interface. These effects have only recently been studied in full depth by *operando* AP-XPS in Sm-doped $CeO_2$ [31]. The authors considered three possible origins for the surface electrical double layer : (i) charge redistribution between the adsorbate and the MIEC surface (ii) local charge variations solely within the electrode, e.g. via accumulation/depletion of oxygen vacancies and/or dopants in the sub surface region and (iii) the intrinsic dipole moment of adsorbates with no contribution from the bulk. They analysed the overpotential dependent adsorbate coverage, Ce oxidation state and surface dipole potential using three different adsorbate chemistries (polar $OH^-$, nonpolar $CO_3^{2-}$, adsorbate free in Ar). Combining these measurements, the intrinsic dipole moments of the adsorbates could be identified as the main contribution to the electrostatic surface bilayer, explaining the observed overpotential dependent surface potential and expanding the understanding of adsorbate chemistry on reaction kinetics.

Finally, another distinctive synchrotron based technique for *operando* studies is anomalous surface XRD, which can provide very valuable 3D structural and chemical information at the atomic level for a given interface. This technique was used for the study of the LSC/YSZ electrode/electrolyte buried interface, as part of a whole cell, under polarization and at different temperatures and oxygen partial pressures [22]. Anomalous SXRD data sets were collected by measuring crystal truncation rods (CTRs), i.e. lines of diffracted intensity in reciprocal space, at the K-edge energies of Y and Zr, as reproduced in Figure 2(f). By choosing these two energies it was possible to distinguish between Y and Zr despite their almost identical atomic form factors, and by combining measurements at several energies, it was possible to obtain composition values with higher accuracy. The Y concentration at the interface was found to vary from Y-rich after deposition, to strongly reduced after being exposed to oxygen at high temperature (approx. 500 °C), as shown in Figure 2(g). Its concentration also changed upon the application of a bias, being enhanced for positive field. Interestingly, yttrium segregation was corroborated by an outward interfacial relaxation of the YSZ metal ion layer, and an increase in point defect concentration was also observed. These interfacial chemical changes are expected to impact the oxygen ion transport at the interface. Thus, disentangling their behavior under operation conditions is key to understand, and ultimately improve, the performance of SOFC devices.

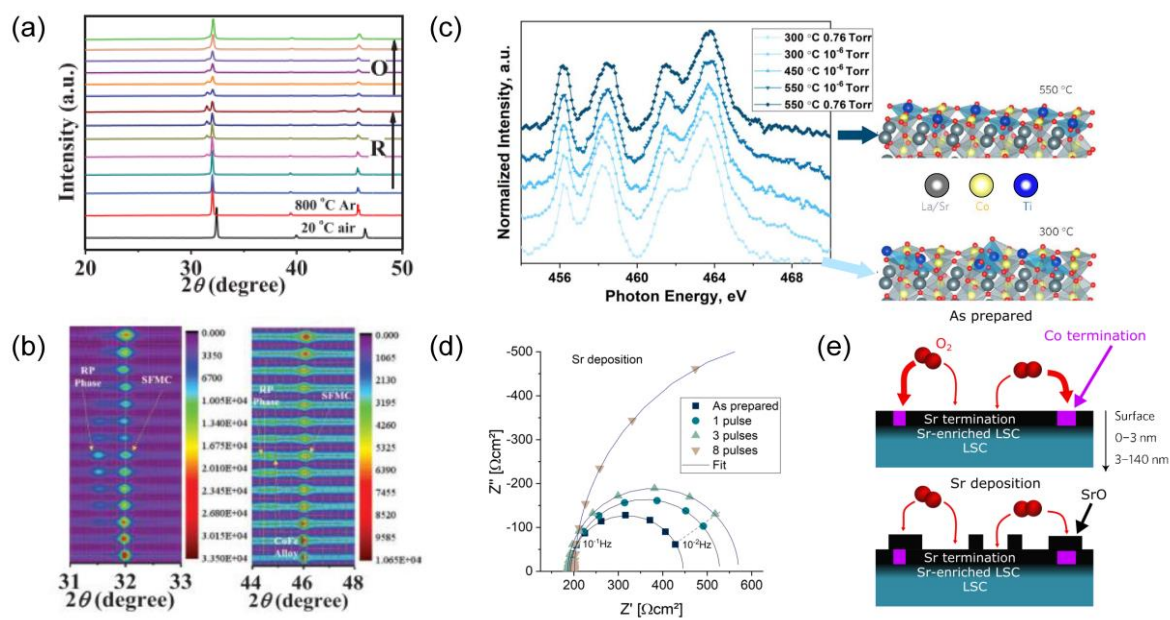

*Figure 3: (a) Reversible phase transition of $Sr_2Fe_{1.35}Mo_{0.45}Co_{0.2}O_{6-\delta}$ and exsolution/dissolution of CoFe nanoparticles from host observed by environmental high temperature XRD at 800°C upon changing the atmosphere from Ar to reducing (R) and oxidising (O) conditions; regions around the SFMC peaks are enlarged in (b) [32]. © 2020 WILEY-VCH. (c) Stabilisation of the LSC surface by cationic Ti decoration on LSC: in situ x-ray absorption spectra of Ti on LSC-Ti15 indicate that above 450°C Ti occupies the perovskite B-site, modifying reducibility and effectively suppressing Sr segregation [33]. Figure kindly provided by B. Yildiz/Q. Lu. (d) Electrochemical behaviour of LSC observed by EIS during in situ decoration with Sr (IPLD) (figure kindly provided by J. Fleig) and (e) schematic of oxygen activity of LSC governed by a few highly active sites (Co termination), being blocked by additional Sr [34].*

## Nanoengineered surfaces via exsolution for enhanced cathode performance

Surface decoration of oxides with catalytically active materials is widely used in the fields of catalysis, energy conversion and storage [35], e.g. via the physical or chemical deposition of metal nanoparticles (NP) on the surface, impregnation and infiltration techniques [36]–[38]. However, agglomeration and growth of these particles during high temperature operation leads to a shrinking active area and degrading catalytic activity [39]. This issue was tackled by an innovative approach: the *in situ* exsolution of catalytic metal NP from the parent material, such as various perovskite structures, under reducing conditions [40]. Doped metallic ions are segregated from the perovskite B-site, with A-site deficiency being a trigger for this exsolution process [41]–[43]. Finely sized exsolved NPs were found to be homogenously distributed and thermally stable, as they are pinned within the parent material. This anchorage avoids agglomeration and thus vastly improves the electrode durability [43]–[46]. While the synthesis and *ex situ* analysis of such materials have gained considerable amount of attraction [47]–[50], direct observations of the exsolution process are still scarce, some of which being highlighted here.

Optical *in situ* characterization techniques provide direct information about the exsolution and growth mechanism of nanoparticles. A pioneer real-time environmental TEM (ETEM) study found that fast exsolution of Co NP from the porous Co doped double perovskite $Pr_{0.5}Ba_{0.5}MnO_x$ occurs after crystal reconstruction at temperatures above 800°C [51]. Recently, an *in situ* TEM analysis showed that Co NP are preferentially extracted from $SrTi_{0.75}Co_{0.25}O_{3-\delta}$ at grain boundaries [52], while anchoring prevents agglomeration. This exsolution was observed under high vacuum and temperatures between 500 and 800°C with a temperature and grain size dependent final particle size and density, respectively. Based on *in situ* observations of the NP nucleation saturation time and size, the authors concluded that the supply of reactants, e.g. Co, limits the exsolution rate. This result has been confirmed by Neagu *et al.* [53]. Their ETEM study of Ni doped $La_{0.8}Ce_{0.1}Ti_{1-x}O_3$ perovskite with high spatial and temporal resolution provided further understanding of the mechanism on an atomic scale. They showed that the isotropical and epitaxial growth of NPs occurs within the surface, while the perovskite lattice forms a volcano shaped socket around each particle. Additionally, the authors found a dependence of the particle shape on the reducing atmosphere, showing a possible routine for nanotailoring exsolved nanostructures.

The effect of the host structure on the exsolution process was exemplarily studied in the Ni doped A-site deficient layered double perovskite $(PrBa)Mn_2Ni_xO_6$ by high temperature neutron powder diffraction [54]. In contrast to previous reports, a comparably slow Ni exsolution at 900°C was observed with a simultaneous phase transition of the host perovskite [51]. A phase transformation of $Sr_2Fe_{1.35}Mo_{0.45}Co_{0.2}O_{6-\delta}$ (SFMC) with concomitant exsolution of CoFe alloy NPs was detected as well by *in situ* XRD measurements [32]. Interestingly, it was reported that iron exsolution from orthorhombic LSF was interrupted around 550°C due to a phase transition, possibly into a fluorite structure, and revived only above 750°C, while rhombohedral LSF powder exhibited a transformation to the cubic structure prior and prerequisite to exsolution [55], [56].

*Lv et al.* showed for the first time *in situ* evidence of a reversible exsolution and dissolution cycle of CoFe nanoparticles in SFMC by changing from reducing to oxidising conditions at 800°C, using *in situ* XRD, TEM and scanning electron microscopy (SEM) [32]. *In situ* X-ray diffraction patterns, showing the reversibility of the materials phase transition from and the exsolution of NPs, triggered by changes in the atmosphere, are reproduced in Figure 3(a) and (b), respectively. The authors reported several remarkable features. The reduction and exsolution of Fe is furthered by the creation of oxygen vacancies through the initial exsolution of facilely reducible Co ions. The local structure remains mostly unaffected after a full cycle of exsolution and reincorporation of CoFe NP, only showing increased roughness. The electrochemical performance is strongly enhanced with increasing number of exsolved particles and excellent redox stability over several cycles was reported.

The effect of $PrO_x$ NPs, exsolved from a perovskite heterostructure (e.g. LSCF), on oxygen activity has been studied by *Chen et al.* by near ambient pressure XPS (NAP-XPS) and XANES measurements [57]. The authors found by NAP-XPS that the $PrO_x$ NPs effectively increase the DOS in the valence band, as compared to pristine LSCF. A temperature dependent decrease of the O K-edge XAS (near edge EXAFS) was assigned to the formation of oxygen vacancies in $PrO_x$. The authors concluded that the enhanced oxygen kinetics in $PrO_x$ coated LSFC originates in a higher number of available oxygen vacancies and electronic states, facilitating the charge transfer.

Tailoring oxide electrodes via the exsolution of NPs has led to promising improvements of their oxygen activity. Further insight in the exsolution process is expected to extend the range of available host and exsolved materials, guiding the development towards advanced electrodes for highly efficient cells, operational at low temperatures.

# Sr segregation: a key degradation issue in perovskite-based cathodes

Mixed ionic and electronic conductive perovskite oxides are very promising cathode materials for intermediate temperature solid oxide fuel cells since they present high stability and a high catalytic activity for the reduction of oxygen (oxygen reduction reaction, ORR). However their efficiency decreases over time due to a number of degradation mechanisms such as interaction with the electrolyte material, chromium and sulphur poisoning, formation of insulating species and coarsening of the microstructure [13]. In addition, the surface of the oxide perovskites undergoes some changes during operation that result in a deterioration of the electrode activity over time. In particular Sr diffusion and segregation from the bulk to the surface together with the volatilization of Sr species have been identified as the main cause behind the loss of activity over time in Sr-substituted, state-of-the-art perovskite materials such as LSC and LSCF [13], [58], [59]. However, despite the many studies performed to date, there are still uncertainties regarding the causes and mechanism behind Sr segregation, the nature of the phases formed and how they impact the activity of the electrode. Recent studies involving *in situ* characterization have contributed to shed light into this subject and have provided important keys towards the improvement of surface stability in oxide perovskite based SOFC cathodes. In this section, results obtained by applying *in situ* near ambient pressure XPS and XAS, SEM, PLD and Raman are detailed.

As shown in the previous sections, the use of *in situ* XPS and XAS studies is very useful since it provides chemical information of the surface of the studied material as it evolves. Tsvetkov, Lu et al. [33] used these approaches to study the effect of modifying the surface of LSC by adding cations that are more and less reducible than Co. The hypothesis behind this study was that the stability of the surface depends on its reducibility. AP-XPS/XAS results showed that different added cations resulted in different oxidation state of Co. In the case of Ti doping, it was shown that the dopant occupies the Co-perovskite site with a stronger cation-O bonding as compared to Co-O (Figure 3(c)), modifying the oxygen vacancy formation energy of LSC. There is indeed a volcano relation between oxygen exchange kinetics and the oxygen vacancy formation enthalpy of the binary oxides of the added cations. This implies that there is a balance between the gain in stability enhancement and the oxygen exchange kinetics, both linked to the amount of oxygen vacancies in the surface, which are in turn reflected in the oxidation state of Co atoms. Conversely to the assumed understanding that surface oxygen vacancies enhance the oxygen reaction kinetics, this study shows that both improved kinetics and stability are obtained when reducing the amount of surface oxygen vacancies. In the best case, when Hf was added to the surface, up to 30 times faster oxygen exchange kinetics were obtained after 54 h at 530 °C in air, when compared to LSC. Wen et al. [60] obtained similar results by performing *in situ* AP-XPS on epitaxial LSC thin films with and without a $ZrO_2$ thin layer deposited by atomic layer deposition (ALD). The $ZrO_2$ coating stabilizes the surface reducing the concentration of oxygen vacancies thanks to a cation exchange between Co and Zr near the surface region, resulting in a more stable and much lower polarization resistance when compared to the uncoated LSC at 550 °C for over 330 h. Finally, Opitz et al. [61] combined the use of *in situ* NAP-XPS with EIS measurements to study Sr segregation in LSC thin films. Their results show the apparition of a La-containing Sr phase that forms initially on top of the surface, which even before covering the whole surface results in a blockage of active Co sites, being the main responsible of the degradation of the electrode activity. Indeed, upon a further increase of temperature, Sr segregation was also observed but it had little effect on the electrode activity. Previous reports had assumed that a SrO monolayer was behind the loss of activity.

The new IPLD method developed by J. Fleig's group (explained in the first section) has also been used to prove the effect of the composition of the surface of oxide perovskites on their catalytic activity. The method consists in coupling the modification of the perovskite surface by the addition of incomplete thin layers by PLD with *in situ* impedance monitoring (see Fig 1 above) [34]. In their study, the surface of LSC was decorated with less than a monolayer of Sr or Co at low temperature to avoid bulk diffusion. Their results show that fractions as low as a 4% of a complete monolayer of SrO have a severe effect on oxygen reduction kinetics, as shown in Figure 3(c), in agreement with the results by Opitz et al. reported above [61]. This would imply that only a few active sites on the surface of LSC would be linked to the high activity of this material. When Co was deposited on the surface a (re-)activation of the activity was observed, thus implying that the active sites are strongly related to the presence of Co at the surface.

Being able to follow the evolution of the morphology of the surface of oxide perovskites when subjected to high temperatures is also very useful. In this line, Niania et al. [62] reported an *in situ* high-temperature environmental scanning electron microscopy (HT-ESEM) study of Sr segregation in LSCF. They used three different atmospheres and compared the evolution of the surface: $O_2$, pure $H_2O$ and ambient lab air. They found that water yields a faster and larger formation of Sr particles. But in all cases 4 common features were identified: i) only Sr particles were observed, ii) the particles form both on the grain and at the grain boundaries, iii) grain boundaries and defects (such as scratches) were preferential nucleation sites, and iv) agglomeration of the different particles happened above a particular temperature for each different atmosphere, a monolayer being formed prior to particle growth. They also identified that common air components (such as $CO_2$, $SO_x$ or $NO_x$) may passivate the surface. Based in their observations the authors propose a mechanism of Sr precipitate formation in several steps: strontium segregation to the surface, monolayer formation, nucleation at grain boundaries or defects, formation of Sr-based particles, and, finally, particle agglomeration.

*In situ* SERS studies have also been used to probe the superior stability of the double perovskite $PrBa_{0.8}Ca_{0.2}Co_2O_{5+\delta}$ (PBCC) as compared to LSC. The study involves the exposure of the two cathode materials to atmospheres containing different amounts of $CO_2$ (from 0 to 10%). *In situ* SERS results showed that carbonates form more readily in LSC than in PBCC. Combining these results with *ex situ* EIS studies, which show a higher stability of PBCC when exposed to $CO_2$, and Density Functional Theory (DFT) calculations, which show that carbonates would be the result of a strong interaction between $CO_2$ and Sr atoms, yield to the conclusion that PBCC presents less Sr segregation to the surface and thus a more stable behaviour. Nevertheless, the study does not explain why while PBCC shows a decrease in performance when 10% $CO_2$ is used despite the SERS results do not show any sign of carbonate formation.

The examples above show how *in situ* studies can shed light on Sr segregation, revealing reaction mechanisms as well as the composition and nature of the surface species involved. Despite this advancement, there are still open questions and controversies remaining, which will need further studies to be clarified. The utilisation of *operando* approaches will certainly contribute by adding more key pieces to this intriguing puzzle.

## Prospects

*In situ* and *operando* studies are key to gain significant insight in the different aspects of SOFC and SOEC cell materials, elementary reaction mechanisms and degradation processes. We have shown a snapshot of the most recent progress in this emerging field, including advanced characterisation tools such as synchrotron based techniques, non-vacuum approaches and high resolution microscopy. The utilization of these tools has already fuelled innovation and led to improvements of materials at stake down to the nanoscale, resulting in enhanced cell performance and durability, enabling novel applications such as full ceramic micro solid oxide fuel cells [63] and extra-terrestrial oxygen production using solid oxide electrolysis cells [64].

However, several limitations and challenges inherent to the *in situ* and *operando* techniques presented in this topical review still remain to date, as will be briefly discussed in the following. While the accessible experimental window for several probes has been successfully extended from vacuum to environmental conditions, ranging from low to ambient pressures, real conditions including ambient air are not yet established. Furthermore, researchers are confronted with the need for nano and sub-nano resolution to further elucidate the relevant chemical and electrochemical processes on an atomic scale. For this purpose fourth-generation extremely brilliant source (EBS) synchrotrons are currently being upgraded to provide microfocussed beamlines with high spatial resolution and multidimensional nano-imaging capabilities (3D, time, chemical, etc.). While TEM provides exceptional spatial resolution down to the atomic level, the sensitivity in conventional imaging modes is significantly reduced for light elements, including oxygen, which plays a major role in SOC electrochemical processes. Hence, the development and use of new TEM methods, such as annular bright field (ABF-STEM) imaging, is necessary for better detection, visualization and quantification of oxygen atoms in the structure. Moreover, as highlighted in the excellent recent review on the topic [65], the probing of oxygen vacancies presents major difficulties due to their dilute concentrations. The techniques available to probe these defects, which play a crucial role in the electrochemical properties of the SOC component materials, are also discussed in their work. As for Raman spectroscopy, we have previously discussed the opportunities of SERS in boosting the sensitivity towards surface species. Howbeit, the fundamental restraint to materials with active Raman modes persist.

Currently, often a single characterization technique is used for each study providing only structural or only chemical information, in some cases not simultaneously combined with the measurements of the material or device functional properties. For future investigations we propose the combination of electrical measurements, Raman spectroscopy, X-ray probes, ellipsometry, (dual) secondary ions mass spectroscopy (SIMS), low energy ion scattering (LEIS) and, of course, of several synchrotron techniques to obtain complementary *in situ* or *operando* information.

Finally, many studies are restricted to the analysis of individual cell components, however the influence of additional interfaces and constituents has to be kept in mind and monitored under working conditions. Therefore, moving to *operando* techniques is required to obtain a complete picture of operation. This need is linked with to the challenge in the design of electrochemical cells suitable for advanced characterisation. While synchrotron studies provide rich information with high temporal resolution, long term effects cannot be studied *in situ* due to the expensiveness and access limitations of large scale facilities. Establishing cheap, lab-based techniques is a prerequisite for manifold long term testing of real SOFCs.

Taking into account the above-mentioned current gaps in the field we anticipate that future advancements in the fundamental understanding and development of SOC materials will be in particular based on:

   I. the gradual evolution of measurement environments from *in situ* to *operando*, approaching real conditions (ambient air, operation temperature, polarization and long term).
   II. the increase in spatial resolution down to the nanometer scale with the advent of high-brilliance synchrotron radiation sources
   III. the sophisticated combination of simultaneous *in situ* and *operando* studies through multimodal approaches for a more holistic understanding of the electrochemical, chemical and physical processes taking place.
   IV. *in situ* characterisation techniques directly linked to the synthesis process, revealing information about "fresh", as-deposited surfaces

Further progress and exploitation of these techniques will continuously stimulate improvements of the employed materials, not only for SOC, but generally in the field of energy systems.

## Acknowledgements

This project has received funding from the European Union's Horizon 2020 research and innovation programme under grant agreement No 824072 (Harvestore).